\documentclass{article}
\usepackage{iclr2021_conference,times}
\usepackage{hyperref}
\usepackage{url}
\usepackage{graphicx}

\usepackage{amsmath,amsfonts,bm}

\def\eqref#1{equation~\ref{#1}}

\def\1{\bm{1}}

\DeclareMathAlphabet{\mathsfit}{\encodingdefault}{\sfdefault}{m}{sl}
\SetMathAlphabet{\mathsfit}{bold}{\encodingdefault}{\sfdefault}{bx}{n}

\usepackage{textcomp}

\newcommand{\comment}[1]{}

\title{Predicting the Binding of SARS-CoV-2 Peptides to the Major Histocompatibility Complex with Recurrent Neural Networks}

\author{Johanna Vielhaben, Markus Wenzel, Eva Weicken \& Nils Strodthoff\\
Department of Artificial Intelligence\\
Fraunhofer Heinrich Hertz Institute\\
Berlin, Germany \\
\texttt{firstname.lastname@hhi.fraunhofer.de}}
\iclrfinalcopy 

\begin{document}
\maketitle

\begin{abstract}
Predicting the binding of viral peptides to the major histocompatibility complex with machine learning can potentially extend the computational immunology toolkit for vaccine development, and serve as a key component in the fight against a pandemic. In this work, we adapt and extend \textit{USMPep}, a recently proposed, conceptually simple prediction algorithm based on recurrent neural networks. Most notably, we combine regressors (binding affinity data) and classifiers (mass spectrometry data) from qualitatively different data sources to obtain a more comprehensive prediction tool. We evaluate the performance on a recently released SARS-CoV-2 dataset with binding stability measurements. \textit{USMPep} not only sets new benchmarks on selected single alleles, but consistently turns out to be among the best-performing methods or, for some metrics, to be even the overall best-performing method for this task.
\end{abstract}

\section{Introduction}
\label{introduction}

Predicting the binding between viral peptides and human proteins from the adaptive immune system using machine learning may serve as a valuable tool to increase the speed of vaccine development in the ongoing SARS-CoV-2-pandemic as well as in future health crises. Accelerated vaccine development supported by computational biology tools may become especially relevant against the background of an evolutionary arms race between viral escape variants and vaccine adaptation until herd immunity can finally be reached. 

Major histocompatibility complex (MHC) molecules encoded by the human leukocyte antigen (HLA) gene complex, play a crucial role in the adaptive immune system \citep{Klein2000, Wieczorek2017}. They induce an immune response by presenting antigen fragments on the cell-surface to immune effector cells \citep{Wieczorek2017,Vyas2008} , and therefore take part in gaining acquired immunity through vaccination. E.g., novel RNA-based vaccines against SARS-CoV-2 enter human cells and elicit the expression of viral spike proteins. They are broken down by the proteasome into antigen peptides which bind to MHC proteins with varying binding affinity. Bound antigen peptides (protein-derived epitopes) are presented by MHC on the cell surface and tie to T-cells that trigger an immune response leading to acquired immunity \citep{Sahin2014}. 

MHC is highly polymorphic such that humans express individual combinations of MHC alleles that bind differently tight to a given peptide, which can affect the potency of an evoked immune response \citep{Winchester2008}. Moreover, there are different MHC classes. MHC class~I molecules are found on almost every nucleated body cell and on platelets at varying densities. They continuously present fragments of proteins produced in the cell -- self or non-self antigens (e.g., viruses) -- to CD8 T cells \citep{Groothuis2005, Shastri2005}. MHC class~II occurs mainly in professional antigen presenting cells of the immune system (e.g., B-lymphocytes) where they present fragments of extracellular ingested pathogens to CD4 T cells \citep{Vyas2008}.

At present, several (e.g., mRNA-based) COVID-19-vaccines make use of the amino acid sequence of the SARS-CoV-2 spike protein, which constitutes about 1/8 of the viral proteome~\citep{Prachar2020}.
Viral escape variants of the spike protein that would degrade into peptides that bind less tight to MHC can be expected to become more prevalent due to evolutionary pressure as a result of widespread vaccine campaigns~\citep{weisblum2020escape}. 
In this case, it might be necessary to leverage selected parts of the remaining 7/8 of the viral proteome for novel vaccine candidates~\citep{Prachar2020, grifoni2020sequence}.
Identifying and increasing the number of immunodominant B- and T-cell epitopes (while excluding those that may even cause adverse effects) is a potential strategy in vaccination design to generate protective immunogenicity \citep{dong2020systematic}.
Multi-epitope vaccines against SARS-CoV-2 might be able to achieve a more precise immune response and to limit the risk of allergic reactions \citep[see][]{kar2020candidate}. 
While full experimental characterization of all potential peptides of several virus variants is slow or might not be feasible at all, prioritization by MHC-peptide binding stability prediction may substantially accelerate the development of a more effective vaccine \citep{Prachar2020, grifoni2020sequence}. This approach may also enable the creation of epitope vaccines targeted against several virus strains at the same time.

A wide range of binding affinity prediction methods based on machine learning has been developed with potential application to vaccine development as well as to personalized cancer immunotherapy. These methods are summarized in a recent comparative review \citep[][]{Zhao2018}; see also \citet{Prachar2020} for a comparison with particular focus on SARS-CoV-2. At this point, it is worth stressing that many of the established methods rely on complicated training procedures with intricate model selection procedures and/or rely on heuristics to identify, e.g., binding regions.

In this work, we evaluate the performance of a novel algorithm for peptide-MHC binding affinity/stability prediction on a recently released dataset with binding stability measurements between SARS-CoV-2 peptides and ten alleles of MHC class~I and one allele of MHC class~II~\citep{Prachar2020}. The algorithm is based on recurrent neural networks and was recently proposed as \textit{USMPep}~\citep{Vielhaben2020}. 
The publication of the dataset by \citet{Prachar2020} also contains a benchmark comparison of about twenty state-of-the-art prediction algorithms (published before 2 March 2020) on these new binding stability measurements. With this contribution, we provide an update for this benchmark by adding the results of an extended version of \textit{USMPep} (which was published on 2 July 2020, i.e. after the `reporting date' of \citet{Prachar2020}).

\section{Materials and Methods}
\label{materials-and-methods}

\paragraph{Datasets \& Targets}
Objective of our work is to predict the binding stability between SARS-CoV-2 peptides and MHC based on the amino acid sequences of the peptides. For this purpose, we train and finally evaluate recurrent neural networks on three different types of lab measurements, involving a peptide of known amino acid sequence and a given MHC allele. Therefore, we distinguish three qualitatively different kinds of targets: During training, we encounter binding affinity (BA) for peptides and mass-spectrometry-eluted (MS) ligands. Whereas the former represents a continuous target (leaving aside qualitative binding affinity labels as provided by \citet{ODonnell2018}), the latter only yield positive (i.e.\ binding) samples, which are typically combined with artificial negative samples in order to be able to train a classifier on this data. Finally, during test time, we aim to predict binding stability (BS), which is also a continuous target, but quantifies the stability of the binding and is hence a more specific measure than binding affinity \citep{Harndahl2012,Jorgensen2014}.
Due to a lack of appropriate training data, we use BA as a proxy target for BS. 

We use a MHC class~I BA dataset provided by \citet{ODonnell2018} and a MS dataset compiled by \citet{Jurtz2017} for model training. For MHC class~II alleles, we train our models on a BA dataset from \citet{Jensen2018} and a MS dataset from \citet{Reynisson2020}. All datasets are based on data retrieved from the Immune Epitope Database \citep{Vita2018}. MS datasets additionally include artificial decoys. We evaluate our tools on the aforementioned BS dataset provided by \citet{Prachar2020}, where the stability measurements are normalized to an allele-specific reference peptide.

\paragraph{Evaluation metrics} We consider the most predominantly used metrics in the field \citep{Zhao2018,Prachar2020}, namely Spearman's $\rho$ and the area under the receiver operating curve (AUCROC) upon framing the task as a classification task using a threshold value of $60\%$ stability. In order to compare the overall performance, we follow \citet{Vielhaben2020} and consider summary metrics, in this case the median due to the small number of alleles under consideration, across alleles.

\paragraph{Model}
We build our approach on \emph{USMPep}, a recently proposed, conceptually simple yet very powerful method \citep{Vielhaben2020}, which is based on a recurrent neural network, in this case with a single-layer long short-term memory (LSTM) architecture. We focus on single-allele models, and precondition the model weights based on an (up to the classification head) identical architecture pretrained with an autoregressive language model objective
\citep{Vielhaben2020}, which generally lead to slight but consistent improvements compared to training from scratch. We consider ensembles of ten individual models for improved stability.

The quantitative and qualitative subsets of the available data let us consider two training objectives: On the log-transformed BA data, we train a regression model (\textit{USMPep\_BA}), as in \citet{Vielhaben2020} using a modified mean squared error loss function that allows to include also qualitative BA measurements~\citep{ODonnell2018}. To leverage the additional, complementary data available through qualitative MS measurements, we train separate classification models using the epitopes identified via MS as well as the artificial negative samples provided in the original MS data using a crossentropy loss.
Finally, we consider combined BA+MS ensembles (\textit{USMPep\_BAMS}) by averaging log-transformed BA and MS predictions, for the first time in the MHC binding prediction literature, to the best of our knowledge. The source code for training and evaluating our models is available at \url{https://github.com/nstrodt/USMPep}. 

\section{Results}
\begin{figure}[h]
\begin{center}
\includegraphics[width=\textwidth]{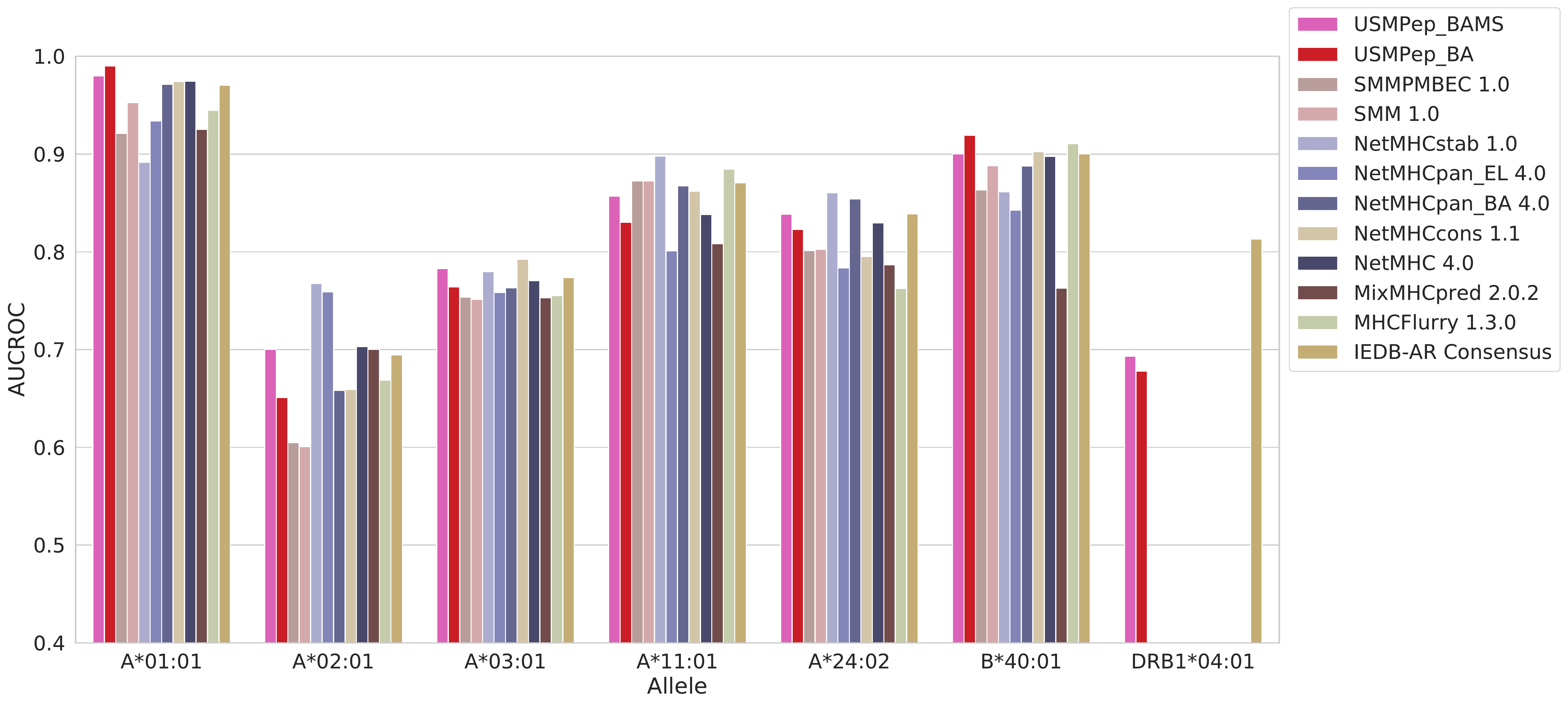}
\end{center}
\caption{Predictive performance on single alleles: Performance of \textit{USMPep} in predicting the binding stability of SARS-CoV-2 peptides and MHC alleles in comparison with state-of-the-art tools reported in Table 2 of \citet{Prachar2020}. The prediction problem was framed as classification task, and the predictive performance was measured using AUCROC as metric. Allele DRB1*04:01 belongs to MHC class~II, all other alleles to class~I.
}
\label{fig-auc}
\end{figure}

Figure~\ref{fig-auc} and Table~\ref{tab-spear-auc-median} compare the performance of \textit{USMPep} to other state-of-the-art methods. We show the performance on single alleles based on AUCROC in Figure~\ref{fig-auc}. 
Both \emph{USMPep}-variants improve the current state-of-the-art for allele A*01:01, the allele with the overall best performance. For B*40:01, \textit{USMPep\_BA} raises the current state-of-the-art to a new level. \textit{USMPep\_BA} is the only tool in the benchmark that is trained on BA data and achieves the highest AUCROC on more than one allele.
While \emph{USMPep} is one of the few tools that provide predictions for the only MHC class~II allele in the test set (DRB1*04:01), its performance on this allele is weaker in comparison to the few other available tools.

Turning to the overall predictive performance in terms of median AUCROC and Spearman's $\rho$ as shown in Table~\ref{tab-spear-auc-median}, \emph{USMPep\_BAMS} turns out to be the overall best-performing method among all tools in terms of Spearman's $\rho$ and show the fourth best performance in terms of AUCROC. 
In particular, the ensembling of five regressors trained on BA data and five classifiers trained on MS data considerably improves the overall performance compared to an ensemble of ten regressors on BA-data alone (\emph{USMPep\_BA}). It is noteworthy that the \emph{USMPep\_BAMS}-ensemble profits from the diversity of the different predictors, e.g.\ a regressor trained on both BA and MS data along the lines of \citep{ODonnell2018} in an ensemble with a classifier trained on MS data yields a weaker performance (Spearman's $\rho$ 0.5260 and AUCROC 0.8097). In summary, these results establish both the original \emph{USMPep\_BA} but in particular the newly proposed \emph{USMPep\_BAMS} as strong predictors for MHC binding stability compared to other state-of-the-art tools.
\begin{table}[t]
\caption{Overall predictive performance: The performance of \textit{USMPep} was assessed with the median Spearman's $\rho$ between predicted binding probability and actual BS across alleles. Besides, the median AUCROC across alleles was evaluated. The results of the state-of-the-art-methods were extracted from Figure 2 of \citet{Prachar2020}. Because numerous tools do not provide predictions for alleles C*01:02, C*07:01 and DRB1*04:01, these were excluded for the median of Spearman's $\rho$ and AUCROC. AUCROC was only evaluated on alleles with more than ten stable binders, which further excludes two remaining HLA-C alleles. The five highest scores are marked in bold for both metrics and are underlined for the best-performing methods.}
\label{tab-spear-auc-median}
\begin{center}
\begin{tabular}{lll}
\multicolumn{1}{c}{\bf Model}  &\multicolumn{1}{c}{\bf{Spearman's $\rho$}}  &\multicolumn{1}{c}{\bf AUCROC}
\\ \hline \\
\textit{USMPep\_BAMS} &            \underline{\textbf{0.56085}} &         \textbf{0.84785} \\
NetMHCstab 1.0    &            \textbf{0.51745} &         \underline{\textbf{0.86080}} \\
NetMHCpan\_BA 4.0  &            \textbf{0.51610} &         \underline{\textbf{0.86080}} \\
IEDB-AR Consensus &            \textbf{0.51440} &         \textbf{0.85470} \\
\textit{USMPep\_BA}      &            \textbf{0.50280} &         0.82660 \\
NetMHC 4.0        &            0.49545 &         0.83385 \\
NetMHCpan\_EL 4.0  &            0.49395 &         0.79235 \\
NetMHCcons 1.1    &            0.49285 &         0.82865 \\
MixMHCpred 2.0.2  &            0.48000 &         0.77485 \\
SMMPMBEC 1.0      &            0.46845 &         0.83235 \\
SMM 1.0           &            0.46540 &         \textbf{0.83760} \\
MHCFlurry 1.3.0   &            0.44265 &         0.82350 \\
\end{tabular}
\end{center}
\end{table}

\section{Summary and Discussion}

We evaluate a novel MHC binding prediction tool on recently published BS measurements involving SARS-CoV-2 peptides. The \textit{USMPep} algorithm is characterized by a conceptually simple architecture and training procedure, can process peptides of arbitrary length and does not rely on further heuristics.  In order to exploit more training data, we adapt and extend the algorithm to consider not only quantitative BA, but also qualitative MS measurements. We find a very high overall performance of \textit{USMPep} in comparison to other state-of the-art methods, and \textit{USMPep} even outperforms all existing methods on selected single alleles. The method can potentially extend the computational immunology toolkit, and help to accelerate vaccine development, and to prevent future epidemics.

Several limits of the work should be considered. Training a model (on BA and MS measurements as proxy) in order to predict the binding of a given peptide to a certain MHC allele can only serve as first step. It neither necessarily implies BS \citep[as pointed out by][]{Prachar2020} nor immunogenicity, nor efficacy, nor safety of a potential (e.g.,\ RNA-based) epitope vaccine derived from the amino acid sequence of the peptide. Moreover, additional BS measurements covering a wider range of MHC alleles appear necessary to realise the full potential of this and other prediction tools; in particular in order to warrant that the global population can profit to the same degree in a fair manner, since MHC allele expression may vary with sex and ethnicity \citep{Schneider-Hohendorf2018, QuinonesParra2014}.

\newpage

\bibliography{usmpep_cov}
\bibliographystyle{iclr2021_conference}

\end{document}